\documentclass[11pt,twoside]{article}

\usepackage{ihep,epsfig,amsfonts,amsmath,amssymb,amsbsy}

\usepackage{graphicx}
\usepackage[dvips]{color}
\usepackage{amsmath}
\definecolor{mycolor}{gray}{.8}
\newcommand{\beeq}{\begin{equation}}
\newcommand{\eneq}{\end{equation}}
\newcommand{\beeqn}{\begin{eqnarray}}
\newcommand{\eneqn}{\end{eqnarray}}

\begin{document}

\begin{center}

{\bf \Large{QUANTUM GRAVITY THROUGH NON-PERTURBATIVE \\[2mm]
RENORMALIZATION GROUP 
and IMPROVED BLACK HOLE}}\\

\vspace*{0.5cm}
{\bf\large Hiroki Emoto}\footnote[1]{dantes11@msn.com} \\

{\it 266-11-303 Nakano-shinden,
Surugaku, Shizuoka 422-8051, Japan}\\

{\it Institute for High Energy Physics, 14220, Protvino, Russia}\\
\end{center}

\vspace*{0.5cm}
\begin{abstract}
Usually, General Relativity(GR) is known to be unrenormalizable 
perturbatively from the viewpoint of quantum field theory. But in the 
modern sense of renormalizability, there still remains the
possibility to 
investigate whether GR is "nonpertubatively" renormalizable or not.
Here I 
review the basics and results in this topic based on the "Effective
Average 
Action" approach which was proposed by M.Reuter, and discuss its
application 
to the balck hole geometry.
\end{abstract}

\section*{Introduction and Motivation}

General Relativity(GR) has been investigated for a long time and
widely 
accepeted as classical theory of gravity, but the quantum structure
of the 
gravity is still not completely understood. 
At present, there are  many candidates 
of quantum gravity. Among them,  in this paper , I introduce the
recent 
progress of one approach by "non-perturvative"  renormalization
group, which 
is the search for UV fixed point that S.Weinberg called as
"asymptotic 
safety"~\cite{Wein}.

If we proceed along the course of traditional quantum field theory,
we meet 
the difficulty of nonrenormalizability because its method depends on
the 
perturbative series expansion of the Newton coupling as is the case
of other 
field theory. On the other hand, there exists the modern viewpoint of 
renormalization theory started by K.G.Wilson~\cite{Wilson}. In the
gravity, 
"asymptotic safety" which was pointed out by Weinberg is essentialy
the same 
idea with Wilsonian approach. In his article, he practiced this idea
in 
($2+\epsilon$) dimension by $\epsilon-$expansion. However this method
does 
not work in higher dimensions. In this direction, the first
analytical 
application of the modern "nonperturbative renormalization group
method" to 
General Relativity was started by M.Reuter~\cite{Reu1}.

This methodology has been developed and showed its efficiency in
other field 
theories recently~\cite{Wette}. The problem is reduced to the flow
equation 
of effective (average) action. We review the basics of this
formulation in 
this aritcle, and discuss about the application to try to see the
quantum 
correction for the black hole.

In the end of this section, I would like to mention  related works on 
this UV fixed point analysis in the discretized numerical simulation, 
Dynamical Triangulation and so on~\cite{Amb}, \cite{Yukawa}. In these
cases  
also, the searches for a fixed point or a phase transition structure
have been 
reported  by many authors. Their connection with continuum theory
which I 
will concern below was also studied in~\cite{Reu-Leu}.

\section{Asymptotic safety}

"Criteria of quantum fundamental theory"

(A) perturbative renormalizability in ordinary sense of quantum field
theory.
       (Gaussian fixed point)

               $\ast$ QED, Yang-Mills ,,,,

(B) perturbative nonrenormalizable but well-defined in
nonperturbative 
sense        (non-Gaussian fixed point). 

               $\ast$ renormalization theory of K.G.Wilson(1974) et
               al

               $\ast$ gravity version , "asymptotic safety"
               (S.Weinberg 
1979)~\cite{Wein}

Define demensionless couplings $g_{i}(\mu)= \mu^{d_{i}}
\hat{g}_{i}(\mu)$

\hspace{2cm}        $d_{i}$: mass dimension of original couplings

\hspace{2cm}        $\mu$ : momentum scale of the renormalization
point.

Then, the partial or total reaction rate $R$ is

\begin{equation}
      R=\mu^{D}f(\frac{E}{\mu}, X, \hat{g}(\mu)). 
\end{equation}

Here $D$ is ordinary mass dimension of $R$, and $E$ is some energy 
characterizing the process, $X$ are all other dimensionless physical 
variables.

we set $\mu = E$,

\begin{equation}
      R=E^{D}f( E, X, g(E)). 
\end{equation}

depends on the behavior of the coupling $g(\mu)$ as $\mu \rightarrow
\infty$. 

Then, we concentrate on the flow equation of dimensionless coupling
of 
$g(\mu)$. 

\section{Formulation of nonperturbative RG approach to quantum\\
gravity}

\subsection{Strategy}
The formulation of this problem is the following~\cite{Reu1}

(1) We construct the "effective average action" of gravity in terms
of 
background field method by introducing suitable infrared cutoff term 
(denoted its momentum scale as "$k$"). We suppress the contribution
of 
infrared momentum which is lower than the cutoff scale in the
integral of 
partition function.

(2) We integrate the mode from the UV scale down to the IR scale
gradually, 
and read out the "flow equation" of effective action.

(3) In the original sence, the theory is defined in the infinite
dimensional 
theory space, infinite number of operators,
\begin{equation}
\{\Lambda, R, ,R^2, 
R_{\mu\nu}R^{\mu\nu},R_{\mu\nu\rho\sigma}R^{\mu\nu\rho\sigma}, R^3,
.....\}. 
\end{equation}

To solve the problem, we "truncate" the theory space into the lower
mass 
dimensional operators.
\begin{equation}
\{\Lambda, R  : \textrm{Einstein-Hilbert truncation}\}. 
\end{equation}

Most simple sketch of the evolution equation of the effective average
acion  
$\Gamma_{k}$ is
\begin{equation}
k\partial_{k}\Gamma_{k}=\frac{1}{2}\text{Tr}[(\Gamma^{(2)}_{k}+
\mathcal{R})^{-1}k\partial_{k}\mathcal{R}_{k}]. 
\end{equation}
Ansatz of "Einstein-Hilbert truncation" is
\begin{equation}
\Gamma_{k}[g,\bar{g}]=\frac{1}{16\pi G_k}\int 
d^{d}x\sqrt{g}\{-R(g)+2\bar{\lambda}_k\}+ 
classical\hspace{0.2cm}gauge\hspace{0.2cm}fixing. 
\end{equation}

\subsection{Quantization based on the background field method}
We consider the fluctuations $h_{\mu\nu}$ around the "classical"
background 
$\bar{g}_{\mu\nu}$ in d-dimensional Euclidean sense.

Classical action $S[\gamma]=S[\bar{g}+h]$ is invariant under the
gauge symmetry
\begin{equation}
\delta\gamma_{\mu\mu}=\mathcal{L}_{v}\gamma_{\mu\nu}=\delta
h_{\mu\nu} = 
v^{\rho}\partial_{\rho}\gamma_{\mu\nu}+
\partial_{\nu}v^{\rho}\gamma_{\rho\nu}+\partial_{\mu}v^{\rho}\gamma_{
\rho\nu}, 
\end{equation}
\begin{equation}
(\gamma_{\mu\nu}=\bar{g}_{\mu\nu}+h_{\mu\nu});
\end{equation}
$\mathcal{L}_{v}$ is the Lie derivative with respect to the
infinitesimal coordinate 
transformation, $v^{\mu}$ is the gauge parameter.

The procedure is the same as usual BRS gauge
fixing,$v^{\mu}\rightarrow 
C^{\mu}$, here $C^{\mu}$ is the Faddeev-Popov ghost, and we introduce 
an auxiliary $B$-field, the gauge fixing function $F[\bar{g},h]$
(this imposes  
the gauge condition $F_{\mu}[\bar{g},h]=0$). Then the BRS
transformations 
are
\begin{equation}
\delta_{B}h_{\mu\nu}=\kappa^{-2}\mathcal{L}_{v}\gamma_{\mu\nu}, 
\end{equation}
\begin{equation}
\delta_{B}C^{\mu}=\kappa^{-2}C^{\nu}\partial_{\nu}C^{\mu}, 
\end{equation}
\begin{equation}
\delta_{B}\bar{C}_{\mu}=B_{\mu}, 
\end{equation}
\begin{equation}
\delta_{B}B_{\mu}=0. 
\end{equation}
Here $\kappa\equiv (32\pi G)^{-1}$. 

The lagrangians of the gauge fixing part and the ghost are
\begin{equation}
\mathcal{L}_{GF}=\kappa
B_{\mu}(F^{\mu}+\frac{\alpha\kappa}{2}B^{\mu}), 
\end{equation}
\begin{equation}
\mathcal{L}_{gh}=-\kappa \bar{C}_{\mu}(\delta_{B}F^{\mu}). 
\end{equation}

Here, we choose
\begin{eqnarray}
F_{\mu}=\sqrt{2}\kappa(\delta^{\rho}_{\mu}\bar{D}^{\sigma}-\frac{1}{2
}\bar{g}^{\rho\sigma}\bar{D}_{\mu})h_{\rho\sigma}. 
\end{eqnarray}

We integrate $B$-field at first. Next, we introduce external sources
as 
$\{t^{\mu\nu},\sigma^{\mu},\bar{\sigma}_{\mu}\}$ which couple to 
$\{h_{\mu\nu},C^{\mu},\bar{C}_{\mu}\}$, and
$\{\beta^{\mu\nu},\tau_{\mu}\}$  
couple to the BRS variations of $\{h_{\mu\nu},C^{\mu}\}$
respectively. These 
lead to the following action for the source part
\begin{eqnarray}
&S_{source}= -\int d^d x \sqrt{\bar{g}}\{t^{\mu\nu}h_{\mu\nu} + 
\bar{\sigma}_{\mu}C^{mu} + \sigma^{\mu}\bar{C}_{\mu}
\hspace{1cm}\nonumber\\
&\mbox{} +\beta^{\mu\nu}\mathcal{L}_{C}(\bar{g}_{\mu\nu}+h_{\mu\nu}) 
+\tau_{\mu}C^{\nu}\partial_{nu}C^{\mu}
\}. 
\end{eqnarray}
Then we get the $k$-dependent connected Green's functions
\begin{eqnarray}
\lefteqn{
\exp \{W_k[t^{\mu\nu},\sigma^{\mu},\bar{\sigma}_{\mu}; 
\beta^{\mu\nu},\tau_{\mu};\bar{g}_{\mu\nu}]\}
}\nonumber\\
&=& \int
\mathcal{D}h_{\mu\nu}\mathcal{D}C^{\mu}\mathcal{D}\bar{C}_{\mu}
\exp
\Big\{-S[\bar{g}+h]-S_{gf}[h;\bar{g}]-S_{gh}[h,C,\bar{C};\bar{g}] 
\nonumber \\
&&\mbox{}\hspace{3cm}-{S_{source}-\Delta_k S[h,C,\bar{C};\bar{g}]}
     \Big\}. 
\end{eqnarray}
Here, the cutoff term $\Delta_k S[h,C,\bar{C};\bar{g}]$ was
introduced, and 
the gauge fixing part and the ghost part are
\begin{equation}
\mbox{} S_{gf}=\frac{1}{2\alpha}\int d^d x 
\sqrt{\bar{g}}\bar{g}^{\mu\nu}F_{\mu}F_{\nu}, 
\end{equation}
\begin{equation}
\mbox{} S_{gh}= -\sqrt{2}\int d^d x 
\sqrt{\bar{g}}\bar{C}_{\mu}\mathcal{M}[g,\bar{g}]^{\mu}_{\nu}C^{\nu}, 
\end{equation}
\begin{equation}
\mathcal{M}[g,\bar{g}]^{\mu}_{\nu}\equiv \delta^{\mu}_{\nu}\bar{D}^2
+ 
\bar{R}^{\mu}_{\nu}. 
\end{equation}

\subsection{The cutoff term and the effective average action}

The infrared cutoff term is
\begin{eqnarray}
&\Delta_k S[h,C,\bar{C}:\bar{g}]=\frac{1}{2}\kappa^2\int d^d x 
\sqrt{\bar{g}}h_{\mu\nu}R^{grav}_k[\bar{g}]^{\mu\nu\rho\sigma}+
\nonumber\\
&+\sqrt{2}\int d^d 
x\sqrt{\bar{g}}\bar{C}_{\mu}R^{gh}_k[\bar{g}]C^{\mu}. 
\end{eqnarray}
Here, the cutoff function $R_k[\bar{g}]=\mathcal{Z}_k k^2 
R^{(0)}(-\bar{D}^2/k^2)$ is the interpolating smooth founction which 
satisfies
\begin{equation}
\lim_{x \to \infty} R^{(0)}(x)=0, 
\end{equation}
\begin{equation}
R^{(0)}(x) = \begin{cases}
                0  \hspace{3cm} (x \to \infty)\\
                1  \hspace{3cm} (x \to  0 )
             \end{cases}\hspace*{-3.5mm}.
\end{equation}
This means that eigenvalues, $p^2$, of operator $-\bar{D}^2$ are
integrated 
out for $p^2\gg k^2$ and suppressed for $p^2\ll k^2$ by the
$k$-dependent mass 
term.
For example, we can take
\begin{equation}
R^{(0)}(x;s) = \frac{sx}{\exp{(sx)}-1)}. 
\end{equation}
Here $s$ is an arbitrary constant.

We introduce the $k$-dependent effective average as follows.

The $k$-dependent expectation values are
\begin{equation}
\bar{h}_{\mu\nu}=\frac{1}{\sqrt{\bar{g}}}\frac{\delta W_k}{\delta 
t^{\mu\nu}},~~~~~~\xi=\frac{1}{\sqrt{\bar{g}}}\frac{\delta
W_k}{\delta 
\bar{\sigma}_{\mu}},~~~~~~\bar{\xi}_{\mu}=\frac{1}{\sqrt{\bar{g}}}
\frac{\delta 
W_k}{\delta t^{\mu\nu}}. 
\end{equation}
Then, the $k$-dependent effective action is
\begin{equation}
\tilde{\Gamma}_k [\bar{h},\xi,\bar{\xi};\beta, \tau; \bar{g}]=\int
d^d x 
\sqrt{\bar{g}}\{t^{\mu\nu}\bar{h}_{\mu\nu}+\bar{\sigma}_{\mu}\xi^{\mu
}+\sigma^{\mu}\bar{\xi}_{mu} 
\} - W_k[t,\sigma,\bar{\sigma};\beta,\tau;\bar{g}]. 
\end{equation}
We define the "effective average action" as
\begin{equation}
\Gamma_k[\bar{h},\xi,\bar{\xi};\beta, \tau; \bar{g}]=\tilde{\Gamma}_k 
[\bar{h},\xi,\bar{\xi};\beta, \tau; \bar{g}] - \Delta_k 
S[h,C,\bar{C}:\bar{g}]. 
\end{equation}
If we write the expected value of the quantum metric 
$g_{\mu\nu}\equiv\bar{g}_{\mu\nu}+h_{\mu\nu}$ , and rewrite
$\Gamma_k$ as

\begin{equation}
\Gamma_k[g_{\mu\nu},\bar{g}_{\mu\nu};\xi^{\mu},\bar{\xi}_{\mu};\beta,
\tau]
\equiv\Gamma_k
[g_{\mu\nu}-\bar{g}_{\mu\nu};\xi^{\mu},\bar{\xi}_{\mu};\beta, 
\tau; \bar{g}_{\mu\nu}]. 
\end{equation}

In this language, the effective average action $\Gamma_k$ reduces to
the 
conventional effective action if we set $\beta, \tau = 0$ and $k \to
0$ ,
\begin{equation}
\Gamma[g_{\mu\nu}]=\lim_{k \to 0}\Gamma_k 
[g_{\mu\nu},g_{\mu\nu};0,0;0,0]. 
\end{equation}
This is invariant under the original symmetry $\delta 
g_{\mu\nu}=\mathcal{L}_v g_{\mu\nu}$. In addition, the $k$-dependent
version 
of this is defined as
\begin{equation}
\bar{\Gamma}_k[g_{\mu\nu}]\equiv\Gamma_k
[g_{\mu\nu},g_{\mu\nu};0,0;0,0]. 
\end{equation}
From the original definition of connected Green's function, ($t\equiv 
\ln{k}$)
\begin{equation}
\partial_t W_k = -\frac{1}{2}\text{Tr}\left[<h \otimes h>(\partial_t 
\hat{R}_k)_{\bar{h}\bar{h}} \right]-\text{Tr}\left[<\bar{C} \otimes 
C>(\partial_t \hat{R}_k)_{\bar{\xi}\xi} \right]
\end{equation}
Here
\begin{equation}
(\hat{R}_k)^{\mu\nu\rho\sigma}_{\bar{h}\bar{h}}=\kappa^2(R^{grav}_k[\
bar{g}])^{\mu\nu\rho\sigma},~~~~(\hat{R}_k)_{\bar{\xi}\xi}=\sqrt{2}R^
{gh}_k 
[\bar{g}]. 
\end{equation}
After some manipulations in terms of the connected 2-point function 
$G_{ij}(x,y)$ and its inverse $\tilde{\Gamma}_k^{ij}(x,y)$,
\begin{eqnarray}
&\partial_t \Gamma_k[\bar{h},\xi,\bar{\xi};\beta, \tau; \bar{g}] = 
\frac{1}{2}\text{Tr}\left[(\Gamma_k^{(2)}+\hat{R}_k)_{\bar{h}\bar{h}}
^{-1}(\partial_t 
\hat{R}_k)_{\bar{h}\bar{h}} \right]-  \nonumber \\
&- 
\frac{1}{2}\text{Tr}\left[\left\{(\Gamma_k^{(2)}+\hat{R}_k)_{\bar{\xi
}\xi}^{-1}- 
(\Gamma_k^{(2)}+\hat{R}_k)_{\xi\bar{\xi}}^{-1}\right\}(\partial_t 
\hat{R}_k)_{\bar{\xi}\xi} \right].  
\end{eqnarray}
On the other hand, $\Gamma_k$ satisfies the integrodifferential
equation
\begin{eqnarray}
&
\exp{\Big\{-\Gamma_k[\bar{h},\xi,\bar{\xi};\beta, \tau; \bar{g}]
\Big\}}          = 
\nonumber\\
&= \int \mathcal{D}h\mathcal{D}C\mathcal{D}\bar{C} +
\exp \Big[-\bar{S}[h,C,\bar{C};\beta,\tau;\bar{g}] + \\
&+\int d^d x
  \big\{(h_{\mu\nu}-\bar{h}_{\mu\nu})\frac{\delta\Gamma_k}{\delta 
\bar{h}_{\mu\nu}} +(C^{\mu}-\xi^{\mu})\frac{\delta\Gamma_k}{\delta 
\xi^{\mu}} +(\bar{C}_{\mu}-\bar{\xi}_{\mu})
\frac{\delta\Gamma_k}{\delta 
\bar{\xi}_{\mu}}
  \big\}
      \Big] \times \nonumber \\
&\times\exp \{ -\Delta_k 
S[h-\bar{h},C-\xi,\bar{C}-\bar{\xi};\bar{g}]\}.  \nonumber 
\end{eqnarray}
Here
\beeq
\tilde{S}\equiv S + S_{gf} + S_{gh} - \int d^d 
x\sqrt{\bar{g}}\{\beta^{\mu\nu}\mathcal{L}_C 
(\bar{g}_{\mu\nu}+h_{\mu\nu})+\tau_{\mu}C^{\nu}\partial_{\nu}C^{\mu}
\}. 
\eneq
The main contribution of this functional integral as $k\to \infty$
comes 
from
\beeq
  (h,C,\bar{C}) \sim (\bar{h},~\xi, ~\bar{\xi}).
\eneq
Then, at the UV cutoff scale, the effective average action as
initial 
condition\beeqn
&\Gamma_{\Lambda}[\bar{h},\xi,\bar{\xi};\beta, \tau; \bar{g}]=
S[\bar{g}+\bar{h}] + S_{gf}[\bar{h};\bar{g}] + 
S_{gh}[\bar{h},\xi,\bar{\xi};\bar{g}] \nonumber\\
&- \int d^d x\sqrt{\bar{g}}\{\beta^{\mu\nu}\mathcal{L}_\xi 
(\bar{g}_{\mu\nu}+\bar{h}_{\mu\nu})+\tau_{\mu}\xi^{\nu}\partial_{\nu}
\xi^{\mu} 
\}.
\eneqn

\subsection{Modified Ward identities and truncations}
The BRS variations come from the cutoff and the source terms.
Assuming the 
invariance of the measure
\beeq
<\delta_{\epsilon} S_{source}+\delta_{\epsilon} \Delta_k S>=0.
\eneq
leads to
\beeqn
&\partial_t \Gamma_k[g,\bar{g}] = 
\frac{1}{2}\text{Tr}\left[\kappa^{-2}(\Gamma_k^{(2)}[g,\bar{g}]+R^{gr
av}_k[\bar{g}])^{-1} 
(\partial_t R^{grav}_k[\bar{g}])      \right] - \nonumber\\
&- \text{Tr}\left[(\mathcal{M}[g,\bar{g}] + R^{gh}_k[\bar{g}])^{-1}
\partial_t R^{gh}_k[\bar{g}] \right].\label{evolution}
\eneqn
Here,we take as truncation
\beeqn
\bar{\Gamma}_k[g_{\mu\nu}] &=& \Gamma_k
[g_{\mu\nu},g_{\mu\nu};0,0;0,0]. 
\nonumber\\
&&= \bar{\Gamma}_k[g] + S_{gf}[g-\bar{g};\bar{g}] + 
\hat{\Gamma}_k[g,\bar{g}]
\eneqn
In the first approximation, $\hat{\Gamma}_k[g,\bar{g}]$ will be
neglected.

\subsection{Einstein-Hilbert truncation}
At the UV scale, we set the initial condition as
\begin{equation}
S=\frac{1}{16\pi \bar{G}} \int d^d x \sqrt{-g}\{-R(g)+2
\bar{\lambda}\}, 
\end{equation}
\begin{equation}
\vspace{1cm}G_k \equiv \bar{G}Z_{N_k}. 
\end{equation}
Here, $G_k\equiv \bar{G}/Z_{N_k}$ is the dimensionful renormalized
Newton 
constant. And
\beeqn
&\Gamma_k[g,\bar{g}]=2\kappa^2 Z_{N_k}\int d^d 
x\sqrt{g}\{-R(g)+2\bar{\lambda_k}\} + \nonumber\\
& +\kappa^2 Z_{N_k}\int d^d 
x\sqrt{\bar{g}}\bar{g}^{\mu\nu}(\mathcal{F}_{\mu}^{\alpha\beta}
g_{\alpha\beta})(\mathcal{F}_{\nu}^{\rho\sigma}g_{\rho\sigma}).\label{
ansatz
}
\eneqn
here we take the gauge parameter as $\alpha= \frac{1}{Z_{N_k}}$.

After evaluating the "Trace" of (\ref{evolution})(R.H.S of the
evolution 
equation),  we will compare the LHS of (\ref{ansatz})
\beeq
\partial_t \Gamma_k[g,g]=2 \kappa^2 \int d^d x
\sqrt{g}\left[-R(g)\partial_t 
Z_{N_k}
+ 2\partial_t(Z_{N_k}\bar{\lambda}_k)\right].
\eneq
that is, we project the solution onto the theory space $\{\sqrt{g}, 
\sqrt{g}R  \}$.

To evaluate the functional integral, we contrive in the following
manner.
The quadratic part of the graviton action is
\beeqn
&\Gamma^{quad}_k[\bar{h};\bar{g}]=Z_{N_k}\kappa^2 \int d^d
x\sqrt{\bar{g}}
\{\frac{1}{2}\hat{h}_{\mu\nu}[-\bar{D}^2 -2\bar{\lambda}_k + 
\bar{R}]\hat{h}^{\mu\nu} - \nonumber\\
&-\left(\frac{d-2}{4d}\right)\phi\left[-\bar{D}^2 - 2\bar{\lambda}_k
+ 
\frac{d-4}{d}\bar{R} \right]\phi + \\
& -\bar{R}_{\mu\nu}\bar{h}^{\nu\rho}\bar{h}^{\mu}_{rho} + 
\bar{R}_{\alpha\beta\nu\mu}\bar{h}^{\beta\nu}\bar{h}^{\alpha\mu} + 
\frac{d-4}{d}\phi \bar{R}_{\mu\nu}\bar{h}^{\mu\nu}   \}. \nonumber
\eneqn
Then, we divide the graviton field into trace $\phi\equiv 
\bar{g}^{\mu\nu}h_{\mu\nu}$ and  traceless $\hat{h}_{\mu\nu}$ part
\beeq
h_{\mu\nu}=\hat{h}_{\mu\nu}+\frac{1}{d}\bar{g}_{\mu\nu}\phi, 
~~~~\bar{g}_{\mu\nu}\hat{h}_{\mu\nu} =0. 
\eneq
For technical reasons, we make use of the concrete
form of 
the background spacetime, maximally symmetric spacetime
\beeq
\bar{R}_{\mu\nu\rho\sigma}=\frac{1}{d(d-1)}(\bar{g}_{\mu\rho}\bar{g}_
{\nu\sigma} 
- \bar{g}_{\mu\sigma}\bar{g}_{\nu\rho})\bar{R},\hspace{1cm} 
\bar{R}_{\mu\nu}=\frac{1}{d}\bar{g}_{\mu\nu}\bar{R}.
\eneq
Then, we get
\beeqn
&\Gamma^{quad}_k[\bar{h};\bar{g}]=\frac{1}{2}Z_{N_k}\kappa^2 \int d^d 
x\sqrt{\bar{g}} \{ \hat{h}_{\mu\nu}[-\bar{D}^2 -2\bar{\lambda}_k +
C_T 
\bar{R}]\hat{h}^{\mu\nu} - \nonumber\\
&-\left(\frac{d-2}{2d}\right)\phi\left[-\bar{D}^2 - 2\bar{\lambda}_k
+
C_S 
\bar{R} \right] \}. 
\eneqn
From the view point of the tensor structure, we take the
renormalization 
factor as
\beeqn
&(\mathcal{Z}^{grav}_k)^{\mu\nu\rho\sigma}=\left[(I- 
P_\phi)^{\mu\nu\rho\sigma} -\frac{d-2}{2d} P^{\mu\nu\rho\sigma}_\phi 
\right]Z_{N_k},  \nonumber\\
&\text{where}~~(P_\phi)_{\mu\nu}~^{\rho\sigma}=\frac{1}{d}\bar{g
}_{\mu\nu}\bar{g}^{\rho\sigma}.
\eneqn
Eventually, the quadratic part of the graviton action  is
\beeqn
&\left(\kappa^{-2}(\Gamma_k^{(2)}[g,g] + 
R^{grav}_k[\bar{g}]\right)_{\hat{h}\hat{h}}  = \nonumber\\
& = Z_{N_k}\left[-\bar{D}^2 + k^2 R^{(0)}(-D^2/k^2)
-2\bar{\lambda}_k
+ C_T R 
\right], 
\eneqn
\beeqn
&\left(\kappa^{-2}(\Gamma_k^{(2)}[g,g] + 
R^{grav}_k[\bar{g}]\right)_{\phi\phi} = \nonumber\\
&= -\frac{d-2}{2d} Z_{N_k}\left[-\bar{D}^2 + k^2 R^{(0)}(-D^2/k^2) 
-2\bar{\lambda}_k + C_S R \right]. 
\eneqn
In our truncation here, we don't take account of the evolution for
ghost 
part ($Z^{gh}_k=1$). Therefore
\beeqn
&-\mathcal{M}+R^{gh}_k = -D^2 + k^2 R^{(0)}(-D^2/k^2) + C_V R
\nonumber\\
&\text{with} ~~~~~~~~~C_V \equiv -\frac{1}{d}.
\eneqn
Then, the RHS of renormalization group equation (denoted
$\mathcal{S}_k(R)$) 
with $\bar{g} = g$
\beeqn
&\mathcal{S}_k (R) = \text{Tr}_T \left[\mathcal{N}(\mathcal{A}+ C_T
R)^{-1} 
\right] +\text{Tr}_S \left[\mathcal{N}(\mathcal{A}+ C_S R)^{-1}
\right]  - 
\nonumber\\
&- 2\text{Tr}_V \left[\mathcal{N}_0(\mathcal{A}_0+ C_V R)^{-1} 
\right]\label{trace}.
\eneqn
with
\begin{equation}
\mathcal{A}\equiv -D^2 + k^2 R^{(0)}(-D^2/k^2) -2\bar{\lambda}_k,  
\end{equation}
\beeqn
&\mathcal{N} \equiv \frac{1}{2 Z_{N_k}} \partial_t \left[ Z_{N_k}k^2 
R^{(0)}(-D^2/k^2)\right] = 
\nonumber\\
&=\left[1-\frac{1}{2}\eta_N (k) \right]k^2 R^{(0)}(-D^2/k^2) + D^2 
R^{(0)'}(-D^2/k^2).
\eneqn
Here $\eta_N(k)$ is an anomalous dimension of $\sqrt{g}R$.
\beeq
\eta_N(k)\equiv -\partial_t \ln Z_{N_k}. 
\eneq
We expand (\ref{trace}) in terms of the curvature $R$ up to the
lowest 
order.
\beeqn
&\mathcal{S}_k (R) = \text{Tr}_T \left[\mathcal{N} \mathcal{A}^{-1}
\right] 
+\text{Tr}_S \left[\mathcal{N} \mathcal{A}^{-1} \right] -
2\text{Tr}_V 
\left[\mathcal{N}_0{\mathcal{A}_0}^{-1}
\right] - \nonumber\\
&-R( C_T \text{Tr}_T \left[\mathcal{N} \mathcal{A}^{-2} \right] + 
C_S\text{Tr}_S \left[\mathcal{N} \mathcal{A}^{-2} \right] - 2 C_V 
\text{Tr}_V \left[\mathcal{N}_0{\mathcal{A}_0}^{-2} \right]). 
\eneqn
We can take advantage of the heat kernel formula on the general
background
\beeq
\text{Tr}\left[ e ^{-is D^2}\right] = \left(\frac{i}{4\pi s}
\right)^{d/2} 
tr(I)\int d^d x \sqrt{g}\left\{1-\frac{1}{6}is R + \mathcal{O}(R^2)
\right\}.
\eneq
to evaluation the general functional of operator $W(-D^2)$. Here
$tr_S(I)=1$, $tr_V (I)=d$, \\ $tr_T (I)=(d-1)(d+2)/2$. Using the
Fourier transform 
$\tilde{W}(s)$
\beeq
\text{Tr}\left[ 
W(-D^2)\right]=\int^{\infty}_{-\infty}ds\tilde{W}(s)\text{Tr}\left[e^
{-is 
D^2}\right].
\eneq
and Mellin transform. Then the functional trace is
\beeqn
&\text{Tr}\left[ 
W(-D^2)\right]=\left(\frac{1}{4\pi}\right)^{\frac{d}{2}}tr(I)
\Big\{Q_{d/2}[W]\int 
d^d x \sqrt{g} + \nonumber\\
&+ \frac{1}{6}Q_{d/2 -1}[W]\int d^d x \sqrt{g}R + \mathcal{O}(R^2)
\Big\}
\eneqn
Here
\beeqn
&Q_0[W]=W(0), \nonumber\\
&Q_n[W]=\frac{1}{\Gamma(n)}\int^{\infty}_{0}dz z^{n-1}W(z). 
\eneqn

\section{The running Newton and cosmological constant}

We get two equations of the coupling flow, by comparing the LHS and
the RHS 
from each coefficient of $\sqrt{g}$ and $\sqrt{g}R$
\beeqn
&\partial_t(Z_{N_k}\bar{\lambda}_k)=\frac{1}{4\kappa^2}\left(\frac{1}
{
4\pi}\right)^{\frac{d}{2}}\Big\{tr_T(I)Q_{d/2}[\mathcal{N}/\mathcal{A
}] + 
\nonumber\\
& +tr_S(I)Q_{d/2}[\mathcal{N}/\mathcal{A}] 
-2tr_V(I)Q_{d/2}[\mathcal{N}_0/\mathcal{A}_0] \Big\},
\eneqn
\beeqn
&\partial_t(Z_{N_k})= 
-\frac{1}{12\kappa^2}\left(\frac{1}{4\pi}\right)^{\frac{d}{2}}\Big\{t
r_T(I)Q_{d/2}[\mathcal{N}/\mathcal{A}]  + 
\nonumber\\
&+ tr_S(I)Q_{d/2}[\mathcal{N}/\mathcal{A}] 
-2tr_V(I)Q_{d/2}[\mathcal{N}_0/\mathcal{A}_0] \Big\}.
\eneqn
From here, we express the function $Q_n[z]$ in terms of some
functions 
$\Phi^p_n(w)$ by the definition, for $n>0$
\beeqn
&\Phi^p_n(w)=\frac{1}{\Gamma(n)}\int^{\infty}_{0}dz 
z^{n-1}\frac{R^{(0)}(z)-zR^{(0)'}(z)}{[z+R^{(0)}(z)+w]}, 
\nonumber\\
&\tilde{\Phi}^p_n= \frac{1}{\Gamma(n)}\int^{\infty}_{0}dz 
z^{n-1}\frac{R^{(0)}(z)}{[z+R^{(0)}(z)+w]}, 
\eneqn
and for $n=0$
\beeq
\Phi^p_0(w)=\tilde{\Phi}^p_0(w)=(1+w)^{-p}.
\eneq
In addition, we use the "dimensionless" renormalized coupling
constant
\beeqn
&g_k\equiv k^{d-2}G_k\equiv Z_{N_k}^{-1}\bar{G}, \nonumber\\
&\lambda_k = k^{-2}\bar{\lambda}_k.
\eneqn
Eventually, the evolutions of the couplings, $g_k$ and $\lambda_k$
are
\begin{equation}
\partial_t g_k = [d-2+\eta_N(k)]g_k \label{evo-gravi},
\end{equation} 
\beeqn
&\partial_t \lambda_k = -(2-\eta_N)\lambda_k 
+\frac{1}{2}g_k\left(\frac{1}{4\pi}\right)^{1-
\frac{d}{2}} \times\nonumber\\
&\times[2d(d+1)\Phi^{1}_{d/2}(-2\lambda_k)-8d\Phi^1_{d/2}(0)-d(d+1)
\tilde{\Phi}^1_{d/2}(-2\lambda_k)]\label{evo-lambda}.
\eneqn
Here the anomalous dimension $\eta_N(k)$ is
\beeq
\eta_N(k)=g_k B_1(\lambda_k) + \eta_N(k)g_k B_2(\lambda_k),
\eneq
and
\beeqn
&B_1(\lambda_k)\equiv 
\frac{1}{3}\left(\frac{1}{4\pi}\right)^{1-\frac{d}{2}}\Big[d(d+1)\Phi
^1_{d/2-1}(-2\lambda_k) 
- 6d(d-1)\Phi^2_{d/2}(-2\lambda_k) - 
\nonumber\\
&-4d\Phi^1_{d/2-1}(0)-24\Phi^2_{d/2}(0)  \Big], \nonumber\\
&B_2(\lambda_k)\equiv 
-\frac{1}{6}\left(\frac{1}{4\pi}\right)^{1-\frac{d}{2}}\left[d(d+1)
\tilde{\Phi}^1_{d/2-1}(-2\lambda_k) 
- 6d(d-1)\tilde{\Phi}^2_{d/2}(-2\lambda_k)   \right]. 
\eneqn
the anomalous dimension is also expressed in terms of $g_k$ and
$\lambda_k$
\beeq
\eta_N=\frac{g_k B_1(\lambda_k)}{1-g_k B_2(\lambda_k)}. 
\eneq

\section{Summary of recent results}
We summarize the main results of the numerical 
computation~\cite{Lau-Reu1}, \cite{Lau-Reu2}. This was reported 
in~\cite{Lau-Reu} concisely. The extended theory space which includes
the 
invariant $R^2$ with a coupling $\beta_k$, that is $\{ \sqrt{g},
\sqrt{g}R, 
\sqrt{g}R^2 \}$,  was also investigated (we call it
"$R^2$-truncation" 
here). In each case, the existence of the nontrivial UV
(non-gaussian) fixed 
point was studied for various ways of the cutoff function. The
following 
points are their main features in 4-dimension.

(1)Universal existence: Non-Gaussian fixed point exists for various
type of 
cutoff operators.

(2)Positive Newton constant for any cutoff operators.

(3)Stability: Non-Gaussian fixed point is UV attractive for any
cutoff.

(4)Scheme and gauge dependence: The values of the critical exponents
or each 
quantity $g_k,~\lambda_k $ at UV fixed point are dependent on the
cutoff 
scheme and the gauge fixing condition. But the value of the product
$g_k 
\lambda_k$ at fixed point seemed to be strongly universal.

In addition, the same investigation was performed for 
$R^2$-truncation~[12]. The UV fixed point exists for all 
admissible cutoff in this case also. This  situation is the same for 
Einstein-Hilbert truncation. At the fixed point, the values of
$\beta_k$ are 
always significantly smaller than $\lambda_k$ or $g_k$. This means
the 
validity of the approximation of Einstein-Hilbert truncation for the 
approximation scheme employed ("Einstein-Hilbert dominance").

In higher dimension of spacetime, the existence of UV fixed point
becomes 
more dependent on the approximation scheme than that in 4~dimensions.

As is always the case with the effective action approach, the
trunacation 
with nonlocal invariants was also studied and interesting possibility
was 
also pointed out~\cite{Reu-Sau}

\section{Quantum  Schwarzschild spacetime and curvature 
singularity}

\subsection{Approximate solution for the running coupling in 4
dimensions}

To investigate the quantum  metric of black hole by the 
renormalization group flow in 4 dimension, we study the solution of
the flow 
equations (\ref{evo-gravi}), (\ref{evo-lambda}) at first~\cite{Reu2}.
We consider the case without cosmological constant ($\lambda_k=0$).
\beeqn
\partial_t g(t) = 2\frac{1-\omega'g}{1-B_2}g, 
\eneqn
here
\beeq
\omega'\equiv \omega + B_2, 
\eneq
and
\beeq
\omega=\frac{4}{\pi}\left(1-\frac{\pi^2}{144}
\right),~~~B_2=\frac{2}{3\pi}. 
\eneq
The fixed point is 
\beeq
\begin{cases}
g_*^{\text{IR}}=0 ~~~~~~~~~~~~~~:\text{Gaussian} \nonumber\\
g_*^{\text{UV}}=\frac{1}{\omega'} \sim 0.71~~~:\text{Non-Gaussian}.
\end{cases}
\eneq
The analytical integration is possible
\beeq
\frac{g}{(1-\omega'g)^{\omega/\omega'}}=\frac{g(k_0)}{(1-\omega'g(k_0
))^{\omega/\omega'}}\left(\frac{k}{k_0}\right)^2. 
\eneq
We use the approximation
\beeq
\frac{\omega'}{\omega} \sim 1.18 \to 1, 
\eneq
which corresponds to the first order approximation in the  anomalous
dimension.

Within this approximation, we obtain 
\beeq
g(k)=\frac{g(k_0)k^2}{\omega g(k_0)k^2 + [1-\omega g(k_0)]k_0^2}\;.
\eneq
In terms of  $G(k)\equiv g(k)/k^2 $ and $G_0\equiv G(k_0)$ one gets:
$(k_0 
\sim 0)$.
\beeq
G(k) = \frac{G_0}{1+\omega G_0 k^2}\label{solution}. 
\eneq
This implies the asymptotic freedom. The Newton coupling $G(k)$
vanishes as 
$k^2 \to \infty$.

\subsection{Cutoff function in the Schwarzschild spacetime}

The flow equations (\ref{evo-gravi}), (\ref{evo-lambda}) and its
solution 
(\ref{solution}) are expressed in terms of momentum $k$. On the other
hand, 
we want to know the information  about the solution  in the coordinate
space, 
since the information of curvature or geometrical quantity should be
written 
in the coordinate space. To get such information, we start from the
approach 
suggested in \cite{Reu2}. This "cutoff identification" program has
been 
studied in many papers. It is a generalization of the
relation 
between the quantum  Coulomb potential(Uehling potential)
and the 
solution for the renormalization group equation of electric charge.
That is 
the substitution
\beeq
k \to \frac{1}{r}\label{flat ident}. 
\eneq
in the solution for the running coupling solution
\beeq
{\Large{e^2(k)}}=\frac{e^2}{1-\frac{e^2}{6\pi^2}\ln{\frac{k}{k_0}}}\;.
\eneq
This manipulation surely restores the well-known behaviour of the
effective 
potential up to constant in the leading order at  small enough 
distances~\cite{Reu2}. We consider a generalization of (\ref{flat
ident})
for 
the curved spacetime  below.

There is no universal meaning in a coordinate itself in curved
spacetime.
One way for the identification in the curved background was proposed in~\cite{Reu2}
\begin{equation}
k(P)=\frac{\xi}{d(P)}\;,  \label{ident} 
\end{equation}
\begin{equation}
d(P)\equiv\int_{\mathcal{C}}\sqrt{|ds^2|}\;, \label{proper}
\end{equation}
here $d(P)$ is the proper distance from the center of black hole to
the 
reference point $P$ which corresponds to the infrared cutoff scale in 
momentum space. $\xi$ is some constant which will be determined
later.
We take up the Schwarzschild spacetime, whose metric 
in the Hilbert gauge 
is
\beeq
ds^2=-\left(1-\frac{2G_0 M}{r} \right)dt^2 + {\left(1-\frac{2G_0
M}{r} 
\right)}^{-1}dr^2 + r^2 d\Omega^2.\\
\eneq
For $r<2G_0 M$, the proper distance (\ref{proper}) is
\beeq
d(r)=2G_0 M\arctan\sqrt{\frac{r}{2G_0 M-r}}-\sqrt{r(2G_0 M-r)}.
\eneq
for $r>2G_0 M$
\beeq
d(r)=\pi G_0 M + 2G_0 M\ln\left(\sqrt{\frac{r}{2G_0
M}}+\sqrt{\frac{r}{2G_0 
M}-1} \right) + \sqrt{r(2G_0 M-r)}, 
\eneq
and asymptotically 
\beeqn
&d(r)= r + \mathcal{O}(\ln r)~~~~\text{for}~~ r \to \infty, 
\nonumber\\
&d(r)=\frac{2}{3}\frac{1}{\sqrt{2G_0 
M}}r^{3/2}+\mathcal{O}(r^{5/2})~~~\text{for}~~ r \to 0. 
\eneqn
Therefore we adopt the following function which interpolates the
behaviours 
at $r=0$ and $r=\infty$
\beeq
d(r)=\left(\frac{r^3}{r+\gamma G_0
M}\right)^{1/2}\label{interpolating}. 
\eneq
Here $\gamma=9/2$.

\subsection{RG-modified Schwarzschild spacetime }
We move to the coordinate space by making use of  
solution~(\ref{solution}) through the identification $k=\xi/d(r)$ and
the 
proper distance function adopted in (\ref{interpolating}). We call
the Schwarzschild spacetime  so modified "the RG-modified 
Schwarzschild 
spacetime".

The relation of the Newton constant between some scale $G(k(r))$  and
the 
experimentally observed (infrared) scale $G_0$ is
\beeq
G(r)=\frac{G_0 d(r)^2}{d(r)^2 + \omega\xi^2 G_0}\;.\label{coordinate
solution}
\eneq
At large distances,
\beeq
G(r)=G_0 - \omega \xi^2 \frac{G_0^2}{r^2} + 
\mathcal{O}\left(\frac{1}{r^3}\right).
\eneq
At this stage, we can adjust the parameter to the results from the
effective 
theory calculations in various papers. For example, in \cite{Ham-Liu}
\beeq
\omega\xi^2 = \frac{118}{15 \pi}\;.
\eneq
Then, the lapse function of RG-corrected spacetime at some distance
scale 
$r$ is\beeq
f(r)=1-\frac{2G(r)M}{r}= 1-\frac{2G_0 Mr^2}{r^3 + \omega\xi^2 G_0
[r+\gamma 
G_0 M]}\;.
\eneq
The character of this spacetime was investigated in detail 
in~\cite{Reu2},

$\ast$ For large masses $M$, quantum effects are negligible; 

and its property depends on the mass of the object. For some critical
mass 
$M_{cr}$; 

$\ast$ If $M > M_{cr}$, there is 2 horizons (similar to 
Reissner-Nordstr$\ddot{\mbox{o}}$m);

$\ast$ If $M < M_{cr}$, there is no horizon;

$\ast$ When $M=M_{cr}$, it corresponds to the "Extremal Black Hole" (
Hawking 
evaporation stops). This seems to be the final state of black hole; 

Then, we study its behaviour near the center of the RG-modified black
hole.

Around the UV fixed point momentum or coordinate dependence of the
Newton 
coupling is
\beeq
G(k)\approx \frac{1}{\omega\ k^2}~~~~~ \Leftrightarrow~~~~~
G(r)\approx 
\frac{d(r)^2}{\omega\xi^2}, 
\eneq
for the Schwarzschild background, the distance function $d(r)$ is 
\beeq
d_{Sch}\propto r^{3/2}  ~~~~\to ~~~~ G_{Sch}\propto r^3. 
\eneq
As $r \to 0$, the metric behaves as a "de-Sitter"-like one. 
\beeq
f(r) = 1 - c r^2 \label{Sch}, ~~~~~~~~~ c~ \text{is  some constant}. 
\eneq
On the other hand, if the spacetime becomes de-Sitter background 
approximately near the origin, its distance function is
\beeq
d_{dS}\propto r  ~~~~\to ~~~~ G_{dS}\propto r^2
\eneq
in this case, as $r \to 0$, the lapse behaves
\beeq
f(r) = 1 - c r + \mathcal{O}(r^2)\label{de-Sitter} , ~~~~~~~~~
c~ \text{is some 
constant}.
\eneq
By the way, for the lapse function $f(r) = 1 - c r^{\nu}$ ($\nu>0$),
the 
curvature invariants are
\begin{equation}
R=c(\nu+1)(\nu+2)r^{\nu-2}, 
\end{equation}
\begin{equation}
R_{\mu\nu\rho\sigma}R^{\mu\nu\rho\sigma}= c^2(\nu^4 - 2\nu^3 + 5\nu^2
+ 
4)r^{2\nu-4}.
\end{equation}
There is a curvature singularity at $r=0$ for $\nu<2$. Then the 
(\ref{de-Sitter}) is singular, while the spacetime (\ref{Sch}) is
regular. 
This shows that the above prescription for the cutoff identification
is
not 
consistent.

\subsection{Improvement of the cutoff identification}

The prescription for the cutoff identification in the previous
subsection 
consists of the following steps. (1) We assume the fixed background 
$g_{\mu\nu}$, then calculate the distance function  $d(r)$.(2) 
Identification with momentum space by $k(P)=\frac{\xi}{d(P)}$.(3)The
metric 
modification $\hat{g}_{\mu\nu}$ from the solution of the coupling
flow 
equation.

To make the above prescription consistent, we propose the next 
procedure. This improvement is essentially the same as in the
recent works  \cite{Reu3}, \cite{Reu4}. More detailed explanation will be 
published~\cite{Hiroki}.

(1) We start from the "unfixed" form of the metric $g_{\mu\nu}(r)$,

          \hspace{5cm}  $f(r) = 1 - \frac{2G(r)M}{r}$\;. 

(2) Then, the distance function is
\beeq
d(r)=\int \sqrt{|ds^2|} = \int^r_0 \sqrt{\Big|\frac{r}{r-2G(r)
M}\Big|}\;.
\eneq

(3) On the other hand, we obtained the solution (\ref{coordinate
solution}) 
of the flow equation already in coordinate space.
\beeq
G(r)=\frac{G_0 d(r)^2}{d(r)^2 + \omega\xi^2 G_0}\;. \nonumber
\eneq
Then, we must solve the above two equations simultaneously.
Differentiating 
both equations in $r$, and equating them we have 
\beeq
\frac{1}{2}\sqrt{\frac{G_0-G(r)}{\tilde{\omega}G_0 
G(r)}}\frac{\tilde{\omega}G_0^2 G^{'}(r)}{(G_0-G(r))^2} = 
\sqrt{\Big|\frac{r}{r-2G(r) M}\Big|}\;.
\eneq
This equation cannot be solved exactly. We have to rely  on a 
numerical 
solution  by computer under suitable boundary condition. Here we only
study the 
behaviour near the origin since whether the central singularity of
the 
classical black hole is regularized or not by quantum gravity is
of general 
interest. The spacetime singularity is generated in quite general
situations in 
classical General Relativity~\cite{Hawking}.

When we are near the center $r\ll 2G(r)M$ then 
\beeq
\frac{1}{4}\tilde{\omega}G_0^3 \left(\frac{dG}{dr}\right)^2 = 
-\frac{r}{r-2G(r)M}G(r)(G_0-G(r))^3. 
\eneq
We search for a consistent solution assuming its asymptotic form as
a series 
expansion around $r=0$: 
\beeq
G(r)=\sum_{n=1}^\infty a_n r^n = a_1 r + a_2 r^2 + a_1 r^3 +\cdots\;.  
\nonumber
\eneq
This assumption implies the asymptotic freedom as $r \to 0$.

We get  up to the third order
\beeq
a_1 = 0~,~~~a_2 = \frac{1}{\tilde{\omega}}~,~~~a_3 = 
\frac{M}{2\tilde{\omega}^2}~,~~~\cdots\;.
\eneq
As a result, the lapse function in this case starts with the linear
term. 
Therefore this RG-corrected spacetime has curvature singularity still
at its 
origin. The consistent scheme of the cutoff identification does not
save the 
situation in terms of the resolution of singularity.

\section{Summary and Discussion}
In this talk, we review the basic concept and the formulation of the 
nonperturbative renormalization group approach to quantum gravity.
The 
recent results in numerical survey to this direction are getting to
increase 
the possibility of "asymptotic safety"  in 4 dimension in spite of
its 
nonrenormalizability in the sense of the perturbative theory.

If   this "asymptotic safety" takes place, the Newton constant shows
asymptotic 
freedom in the high energy region, which will correspond to the weak 
coupling between the gravity and the other matter field probably.
Then this 
position-dependent Newton constant will modify our concepts of
spacetime in 
the classical theory.
There are many works which studied the effects of the
position-dependent 
Newton constant in the cosmological problems. The crucial point to
discuss 
the phenomenological impacts is the way in which we interpret the
information 
on the coupling flow solution for momentum space in the coordinate
space. We 
discussed  the quantum modified  Schwarzschild spacetime through
the 
renormalization group by the cutoff identification method which was 
introduced in \cite{Reu2} and tried improving it. This improvement is 
essentially the same concept as the "consistent cutoff
identification"
which was 
proposed already in \cite{Reu3}, \cite{Reu4}. Here we concentrated on
the 
possibility of curvature singularity at the origin in this method.
Whether 
the central singularity of black hole is regularized or not is
of general 
interest in this region. As a result, the curvature singularity still
exists 
at the center of this RG-corrected spacetime also by the improving 
identification method.

Although this identification method shows the interesting effect in
the 
cosomological phenomenology (see for example~\cite{Reu3},\cite{Reu4}),
still 
we could not find the regularization mechanism for the black hole 
singularity or some prescriptions to describe the quantum spacetime
by 
regular geometry. More close argument will be reported
eslewhere~\cite{Hiroki}.

\vspace*{0.4cm}
{\bf\Large{Acknowledgement}}

\vspace*{0.4cm}
I would like to thank A.A.Logunov, V.A.Petrov, A.P.Samokhin for
hearty 
invitation and helpful discussion in the theoretical division of 
IHEP(Protvino) and this international workshop. This work was based
on the 
scientific grant from JINR(Dubna). In addition, I am grateful to
T.Kubota 
for his encouraging discussion in the early stage of this study at
Osaka 
University.

\def\jnl#1#2#3#4{{#1}{\bf #2} (#4) #3}

\def\Zphys{{\em Z.\ Phys.} }
\def\jssc{{\em J.\ Solid State Chem.\ }}
\def\jpsJ{{\em J.\ Phys.\ Soc.\ Japan }}
\def\ptps{{\em Prog.\ Theoret.\ Phys.\ Suppl.\ }}
\def\PTP{{\em Prog.\ Theoret.\ Phys.\  }}

\def\JMP{{\em J. Math.\ Phys.} }
\def\NPB{{\em Nucl.\ Phys.} B}
\def\NP{{\em Nucl.\ Phys.} }
\def\PLB{{\em Phys.\ Lett.} B}
\def\PL{{\em Phys.\ Lett.} }
\def\PRL{\em Phys.\ Rev.\ Lett. }
\def\PRB{{\em Phys.\ Rev.} B}
\def\PRD{{\em Phys.\ Rev.} D}
\def\PRe{{\em Phys.\ Rep.} }
\def\AP{{\em Ann.\ Phys.\ (N.Y.)} }
\def\RMP{{\em Rev.\ Mod.\ Phys.} }
\def\ZPC{{\em Z.\ Phys.} C}
\def\SCI{\em Science}
\def\CMP{\em Comm.\ Math.\ Phys. }
\def\MPLA{{\em Mod.\ Phys.\ Lett.} A}
\def\IJMPB{{\em Int.\ J.\ Mod.\ Phys.} B}
\def\PR{{\em Phys.\ Rev.} }
\def\cmp{{\em Com.\ Math.\ Phys.}}
\def\JPA{{\em J.\  Phys.} A}
\def\CQG{\em Class.\ Quant.\ Grav. }
\def\ATMP{{\em Adv.\ Theoret.\ Math.\ Phys.} }
\def\ibid{{\em ibid.} }
\def\Poi{{\em Ann.\ Inst.\ H.\ Poincare}}

\vspace{1cm}

\leftline{\bf\Large{References}}

\renewenvironment{thebibliography}[1]
        {\begin{list}{[$\,$\arabic{enumi}$\,$]}  
        {\usecounter{enumi}\setlength{\parsep}{0pt}
         \setlength{\itemsep}{0pt}
         \renewcommand{\baselinestretch}{1.2}
         \settowidth
        {\labelwidth}{#1 ~ ~}\sloppy}}{\end{list}}


\begin{thebibliography}{9}
\bibitem{Wein}
S.Weinberg, \textit{General Relativity, an Einstein Centenary
Survey}, 
S.W.Hawking, W.Israel(Eds), Cambridge University Press, 790 (1979)
\bibitem{Wilson}
K.G.Wilson, J.Kogut, \jnl{\PRe}{12}{75}{1974}
\bibitem{Reu-Leu}
O.Lauscher. M.Reuter, hep-th/0508202
\bibitem{Amb}
J.Ambj$\phi$rn, J.Jurkiewicz. R.Loll , hep-th/0505133, hep-th/0505154
\bibitem{Sieg}
W.Siegel, hep-th/0309093
\bibitem{Reu1}
M.Reuter  \jnl{\PRD}{57}{971}{1998}, hep-th/9605030
\bibitem{Wette}
J.Berges, N.Tetradis, C.Wetterich, \jnl{\PRe}{363}{223}{2002}, 
hep-ph/0005122
\bibitem{Ward}
B.F.L.Ward  \jnl{\MPLA}{17}{2371}{2002}, hep-ph/0204102
\bibitem{Hamada}
K.Hamada \jnl{\PTP}{105}{673}{2001},  hep-th/0012053
\bibitem{Yukawa}
H.S.Egawa, S.Horata, T.Yukawa \jnl{\PTP}{108}{1171}{2003}
,hep-lat/0309047
\bibitem{Lau-Reu1}
O.Lauscher, M.Reuter, \jnl{\PRD}{65}{025013}{2002}, hep-th/0108040
\bibitem{Lau-Reu}
O.Lauscher, M.Reuter, \jnl{\CQG}{19}{483}{2002}, hep-th/0110021
\bibitem{Lau-Reu2}
O.Lauscher, M.Reuter, hep-th/0205062
\bibitem{Reu-Sau}
M.Reuter, F.Saueressig, \jnl{\PRD}{56}{125001}{2002}, hep-th/0206145
\bibitem{Reu2}
A.Bonanno, M.Reuter \jnl{\PRD}{62}{043008}{2000}, hep-th/0002196
\bibitem{Ham-Liu}
H.W.Hamber, S.Liu, \jnl{\PLB}{357}{51}{1995}
\bibitem{Reu3}
M.Reuter, H.Weyer ,hep-th/0311196
\bibitem{Reu4}
M.Reuter, H.Weyer, hep-th/0410117
\bibitem{Don}
J.F.Donoghue \jnl{\PRD}{50}{3874}{1994}
\bibitem{Hawking}
S.W.Hawking, G.F.R.Ellis, \textit{The large scale structure of
space-time}, 
Cambridge monographs on mathematical physics, Cambridge university
press 
,1973
\bibitem{Hiroki}
H.Emoto, hep-th/0511075
\end{thebibliography}
\end{document}